\documentclass[12pt,epsfig]{article}
\usepackage{geometry}                
\usepackage{graphicx}
\usepackage{amssymb}
\usepackage{epstopdf}
\usepackage{color}
\begin{document}


\renewcommand{\thesection}{\arabic{section}}
\renewcommand{\theequation}{\arabic{equation}}
\renewcommand {\c}  {\'{c}}
\newcommand {\cc} {\v{c}}
\newcommand {\s}  {\v{s}}
\newcommand {\CC} {\v{C}}
\newcommand {\C}  {\'{C}}
\newcommand {\Z}  {\v{Z}}
\newcommand{\pv}[1]{{-  \hspace {-4.0mm} #1}}

\baselineskip=24pt

\def\beqra{\begin{eqnarray}} \def\eeqra{\end{eqnarray}}
\def\beqast{\begin{eqnarray*}} \def\eeqast{\end{eqnarray*}}
\def\beq{\begin{equation}}      \def\eeq{\end{equation}}
\def\be{\begin{enumerate}}   \def\ee{\end{enumerate}}

\def\gam{\gamma}
\def\Gam{\Gamma}
\def\la{\lambda}
\def\eps{\epsilon}
\def\La{\Lambda}
\def\si{{\rm si}}
\def\Si{\Sigma}
\def\al{\alpha}
\def\Th{\Theta}
\def\th{\theta}
\def\tnu{\tilde\nu}
\def\vphi{\varphi}
\def\del{\delta}
\def\Del{\Delta}
\def\ab{\alpha\beta}
\def\om{\omega}
\def\Om{\Omega}
\def\mn{\mu\nu}
\def\mun{^{\mu}{}_{\nu}}
\def\kap{\kappa}
\def\rsi{\rho\sigma}
\def\beal{\beta\alpha}
\def\til{\tilde}
\def\rta{\rightarrow}
\def\eqv{\equiv}
\def\nab{\nabla}
\def\pa{\partial}
\def\sit{\tilde\sigma}
\def\ul{\underline}
\def\indt{\parindent2.5em}
\def\nd{\noindent}
\def\rsi{\rho\sigma}
\def\beal{\beta\alpha}
\def\caa{{\cal A}}
\def\cb{{\cal B}}
\def\cac{{\cal C}}
\def\cd{{\cal D}}
\def\ce{{\cal E}}
\def\cf{{\cal F}}
\def\cg{{\cal G}}
\def\cah{{\cal H}}
\def\ci{{\rm ci}}
\def\cj{{\cal{J}}}
\def\ck{{\cal K}}
\def\cl{{\cal L}}
\def\cm{{\cal M}}
\def\cn{{\cal N}}
\def\cO{{\cal O}}
\def\cp{{\cal P}}
\def\car{{\cal R}}
\def\cs{{\cal S}}
\def\ct{{\cal{T}}}
\def\cu{{\cal{U}}}
\def\cv{{\cal{V}}}
\def\cw{{\cal{W}}}
\def\cx{{\cal{X}}}
\def\cy{{\cal{Y}}}
\def\cz{{\cal{Z}}}
\def\asymptotic{{_{\stackrel{\displaystyle\longrightarrow}
{x\rightarrow\pm\infty}}\,\, }} 
\def\asymptext{\raisebox{.6ex}{${_{\stackrel{\displaystyle\longrightarrow}
{x\rightarrow\pm\infty}}\,\, }$}} 
\def\epsilim{{_{\textstyle{\rm lim}}\atop
_{~~~\epsilon\rightarrow 0+}\,\, }} 
\def\omegalim{{_{\textstyle{\rm lim}}\atop
_{~~~\om^2\rightarrow 0+}\,\, }} 
\def\xlimp{{_{\textstyle{\rm lim}}\atop
_{~~x\rightarrow \infty}\,\, }} 
\def\xlimm{{_{\textstyle{\rm lim}}\atop
_{~~~x\rightarrow -\infty}\,\, }} 
\def\asymptoticp{{_{\stackrel{\displaystyle\longrightarrow}
{x\rightarrow +\infty}}\,\, }} 
\def\asymptoticm{{_{\stackrel{\displaystyle\longrightarrow}
{x\rightarrow -\infty}}\,\, }} 

\def\raisenot{\raise .5mm\hbox{/}}
\def\nota{\ \hbox{{$a$}\kern-.49em\hbox{/}}}
\def\notA{\hbox{{$A$}\kern-.54em\hbox{\raisenot}}}
\def\notb{\ \hbox{{$b$}\kern-.47em\hbox{/}}}
\def\notB{\ \hbox{{$B$}\kern-.60em\hbox{\raisenot}}}
\def\notc{\ \hbox{{$c$}\kern-.45em\hbox{/}}}
\def\notd{\ \hbox{{$d$}\kern-.53em\hbox{/}}}
\def\notbd{\ \hbox{{$D$}\kern-.61em\hbox{\raisenot}}} 
\def\note{\ \hbox{{$e$}\kern-.47em\hbox{/}}}
\def\notk{\ \hbox{{$k$}\kern-.51em\hbox{/}}}
\def\notp{\ \hbox{{$p$}\kern-.43em\hbox{/}}}
\def\notq{\ \hbox{{$q$}\kern-.47em\hbox{/}}}
\def\notW{\ \hbox{{$W$}\kern-.75em\hbox{\raisenot}}}
\def\notz{\ \hbox{{$Z$}\kern-.61em\hbox{\raisenot}}}
\def\notpa{\hbox{{$\partial$}\kern-.54em\hbox{\raisenot}}}
\def\fo{\hbox{{1}\kern-.25em\hbox{l}}}  
\def\rf#1{$^{#1}$}
\def\bx{\Box}
\def\tr{{\rm Tr}}
\def\rmtr{{\rm tr}}
\def\dgg{\dagger}
\def\lag{\langle}
\def\rag{\rangle}
\def\bmid{\big|}
\def\vlap{\overrightarrow{\La p}} 
\def\lrta{\longrightarrow} \def\lrar{\raisebox{.8ex}{$\longrightarrow$}}
\def\ON{{\cal O}(N)}
\def\UN{{\cal U}(N)}
\def\bdPh{\mbox{\boldmath{$\dot{\!\Phi}$}}}
\def\bPh{\mbox{\boldmath{$\Phi$}}}
\def\bPhs{\bPh^2}
\def\sef{S_{eff}[\sigma,\pi]}
\def\sigx{\sigma(x)}
\def\pix{\pi(x)}
\def\bph{\mbox{\boldmath{$\phi$}}}
\def\bphs{\bph^2}
\def\ex{\BM{x}}
\def\exs{\ex^2}
\def\xdot{\dot{\!\ex}}
\def\y{\BM{y}}
\def\ys{\y^2}
\def\ydot{\dot{\!\y}}
\def\pat{\pa_t}
\def\pax{\pa_x}
\def\hp{{\pi\over 2}}
\def\sign{{\rm sign}\,}
\def\bv{{\bf v}}




\begin{center}
{\bf  Effective Non-Hermitian Hamiltonians for Studying Resonance Statistics in Open Disordered Systems }\\
\bigskip
Joshua  Feinberg{ \footnote{e-mail: joshua@physics.technion.ac.il}}\\
Department of Physics, University of Haifa at Oranim, Tivon 36006,
Israel, \\and \\
Department of Physics, Technion-Israel Inst. of Technology,
Haifa 32000, Israel\\
\bigskip

\end{center}
\setcounter{page}{1}
\bigskip


\begin{minipage}{5.8in}

{\abstract~~~
We briefly discuss construction of energy-dependent effective non-hermitian hamiltonians for studying resonances in open disordered systems . }

\end{minipage}

\bigskip

PACS number(s): 03.65.Yz, 03.65.Nk, 72.15.Rn \\
\bigskip
\bigskip
Keywords: resonances, spectral determinant, disordered systems, average density of resonances

\newpage



\section{Introduction}
\label{section1}
\setcounter{footnote}{0}

Open systems typically give rise to resonances. A resonance is a long-living quasi-stationary state, which eventually decays into the continuum. 
Physically, it may be thought of as a particle, initially trapped inside the system, which eventually escapes to infinity. 

One common approach to studying resonances is based on the analytic properties of the scattering matrix $S({\cal E})$ in the complex energy plane. 
Resonances correspond to poles 
\beq\label{poles}
{\cal E}_n = E_n - \frac{i}{2}\Gamma_n
\eeq 
of $S({\cal E})$ on the non-physical sheet\cite{LL,BPZ}. 
In an alternative equivalent approach, which we shall follow here, one solves the Schr\"odinger equation subjected to the boundary condition of purely outgoing wave outside the range of the potential. 
This boundary condition, which describes a process in which a particle is ejected from the system,  renders the problem non-Hermitian. The Schr\"odinger equation with this boundary condition leads to complex eigenvalues ${\cal E}_n$ which correspond to resonances \cite{LL,BPZ}. For a recent lucid discussion of resonances in quantum systems, with particular emphasis on the latter approach, see \cite{hatano,hf}.  

The outgoing-wave approach leads, in a natural way, to non-Hermitian effective hamiltonians ,whose complex eigenvalues are the resonances of the studied system \cite{wied,rotter,fyod-som}.  Such effective hamiltonians are very useful for studying resonances in scattering theory, including scattering in chaotic and disordered systems\cite{datta,kottos, Stock, Dittes, Pierre, ks1,ks2}. 

There are many examples of resonances in atomic and nuclear physics. Recently, there has been considerable interest in resonances which arise in chaotic and disordered systems. See \cite{kottos} for a recent review.  One of the main goals in these studies is computation of the distribution $P(\Gamma)$ of resonance widths. 
There is ample amount of work on computing $P(\Gamma)$ in one-dimensional disordered chains\cite{ks1,ks2,terraneo,texier,pinheiro,tit-fyod,weiss}. 
Numerical results presented in some of these works indicate that $P(\Gamma)\sim \Gamma^{-\gamma}$ in a large range of values of $\Gamma$, where the exponent
$\gamma$ is very close to $1$.

A more general quantity than $P(\Gamma)$ is the\footnote{In order to avoid cluttering of our formulas, we do not use the resonance width $\Gamma_n$ in (\ref{poles}) as an argument of $\rho$, but rather $y=-\Gamma_n/2$.}
density of resonances ({\bf DOR}) 
\beq\label{dor}
\rho (x,y)  = \sum_{n} \delta (x-{\rm Re} {\cal E}_n)\delta (y-{\rm Im} {\cal E}_n).
\eeq
It is widely believed that the  averaged DOR in the complex plane contains information about the
Anderson transition \cite{kottos, pinheiro, weiss, Kot3}. This expectation \footnote{I learned this argument about the expected scaling behavior of the DOR from B. Shapiro.} is based on an analogy with Thouless' arguments concerning the sensitivity of
eigenstates to the boundary conditions in Hermitian localization theory \cite{Thouless,gang4}. Indeed, the coupling of the
disordered system to the external world plays in our case a role similar to changing the boundary conditions in Thouless' picture.
Namely, the width of a typical resonance in the insulating regime
should be exponentially small, $\Gamma_{typ}\sim \exp -L/\xi(E)$ ($L$ being the size of the system), whereas in the metallic regime
the typical width is $\Gamma_{typ}\sim {\cal D}/L^2$,  namely, the inverse Thouless time scale (${\cal D}$ is the diffusion coefficient in the disordered metal). Thus, $\Gamma_{typ}$, measured in units of
level spacing $\Delta$, is analogous to the Thouless conductance. This picture was already pursued numerically in \cite{Kot3}.

The continuum limit of the disordered chain was studied in \cite{mumbai}. For simplicity, a chain opened only at one end was studied. 
The spectral determinant for the problem was derived, and the averaged DOR was expressed in terms of a certain integral over the solution of a certain  singular two-dimensional Fokker-Planck equation. (That Fokker-Planck equation determined  the probability distribution of the logarithmic derivative of the outgoing wave at the open end of the chain.)

The present work was motivated in part by \cite{ks1,ks2}. In particular, an analytical approach was developed in \cite{ks2} for studying resonances, which is based on counting poles of the resolvent of the non-Hermitian tight-binding effective hamiltonian of the open chain. In the case of a semi-infinite disordered chain, coupled to a semi-infinite perfect lead, these authors have derived an exact integral representation for the DOR, valid for arbitrary disorder and chain-lead coupling strength. In the limit of weak chain-lead coupling (in which resonances are typically narrow) they were able to rigorously derive a universal scaling formula for the DOR, valid for any degree of disorder and everywhere inside  the unperturbed energy band of the closed chain. The $1/\Gamma$ behavior of the DOR follows from that formula. 

In this paper we shall review and explain how to construct energy dependent non-hermitian hamiltonians for studying resonance statistics in open systems. While many (but by no means all) of the results presented in this paper are known, we believe our presentation offers a somewhat fresh look at these issues. Upon elimination of the leads, one can reformulate the problem in terms of an effective non-hermitian hamiltonian, which depends only on the degrees of freedom of the disordered system. In this effective description, the outgoing-wave boundary condition in the original system is translated into a local non-hermitian, energy dependent boundary condition at the contact points (or more generally, contact regions)  of the system and the leads. 

This paper is organized as follows. In Section 2 we discuss resonances in a generic quantum system coupled to the external world by  a single one-dimensional lead 
( a single channel lead). We derive a general expression for the DOR in terms of an appropriate diagonal matrix element of the resolvent of the original closed system. From this expression, we derive an integral representation for the averaged DOR of the disordered system. In Section 3 we specialize to the case of an open one dimensional disordered chain, derive the corresponding effective hamiltonian, and  obtain its continuum limit. The resulting continuum effective non-hermitian hamiltonian differs from the hermitian one of the closed system by a complex energy dependent boundary condition. The structure revealed in this way is quite generic, and we conclude in Section 4 by mentioning similar continuum effective hamiltonians for higher dimensional systems. 

\section{Resonances in a System Connected to a Single Perfect Lead}
\begin{figure}[htb]
\centerline{\includegraphics[scale=0.42]{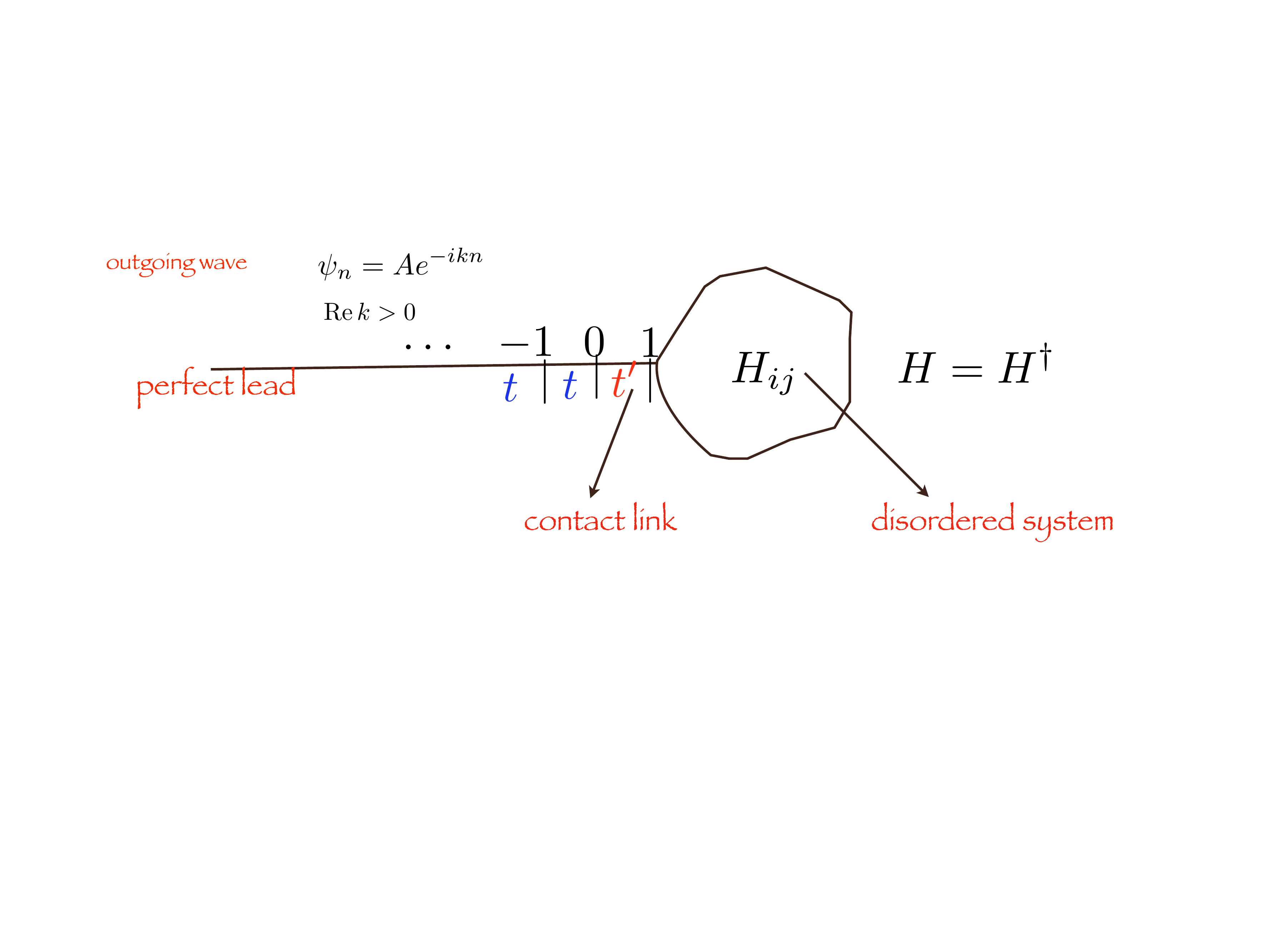}}
\caption{ A disordered system is opened up and coupled to the external world by a perfect lead, 
stretched along the negative $x$-axis. The lead is modeled by a tight-binding hamiltonian, with nearest neighbor hopping amplitude $t$. The $<01>$ link connects the lead and the system, with corresponding hopping amplitude $t'$. The Schr\"odinger equation is to be solved with outgoing wave boundary condition. In this figure we have set the lattice spacing $a=1$. See the text for more details.}
\label{lead}
\end{figure}

In order to keep the discussion as simple as possible, let us consider a quantum system connected to a single perfect semi-infinite one-dimensional lead,
which lies along the negative $x$-axis. This construction is described in Figure \ref{lead}. 
We shall model the lead by means of a tight-binding hopping hamiltonian, with nearest-neighbor hopping amplitude $t$.  
The sites on the lead lie at the points $x_n = n a,\, n= 0,-1,-2,\ldots$, $a$ being the lattice spacing. Let us assume that the (closed) quantum system has an $N$-dimensional  state space, and that it is described by an $N\times N$ {\em hermitian} matrix $H_{ij}$, $H=H^\dagger$. We further assume that 
the system lives on some graph with $N$ nodes, and that  $H_{ij}$ is the matrix element connecting site $i$ to site $j$ (the link $<ij>$ is directed, of course). 
Let us now connect the lead's end $n=0$ to some site in the system, which with no loss of generality we pick to be site $i=1$. The hopping amplitude along the buffer link
$<01>$ is taken to be $t'$, which need not be equal to $t$. In particular, $t'=0$ corresponds to a closed hermitian system. Disorder is modeled by some probability distribution for the matrix $H$, which means, in general, both random hopping and random site energies on the graph. 

The Schr\"odinger equation for this system is therefore 
\beqra\label{schrodinger}
-t(\psi_{n+1} + \psi_{n-1}) &=& z \psi_n\,,\quad n\leq -1\nonumber\\
-t\psi_{-1} -t'\psi_1 &=&  z\psi_0\,,\quad n=0\nonumber\\
-t'\psi_0 + (H\vec\psi)_1 &=& z\psi_1\,\quad n=1\nonumber\\
(H\vec\psi)_n &=& z\psi_n\,,\quad n\geq 2\,.
\eeqra
Here we lumped the wave-function amplitudes inside the system into an $N$-dimensional vector $\vec\psi$, and $z$ is the complex eigenvalue. 
Imposing outgoing-wave boundary condition in the perfect lead means that 
\beq\label{outgoingbc}
\psi_ n  = Ae^{-ikna}\,,\quad n\leq 0\,,
\eeq
where the wave-vector $k$ must be restricted to the right half of the fundamental Brillouine zone, namely, 
\beq\label{leftmoving}
0\leq {\rm Re}\, ka \leq \pi\,,
\eeq
as the wave propagates freely to the left, into the lead. (This choice has the obvious continuum limit ${\rm Re}\, k\geq 0$ describing free propagation to the left.) 

We shall now eliminate the lead entirely from (\ref{schrodinger}), following the idea presented in \cite{datta}. To this end we first substitute (\ref{outgoingbc}) in the first equation 
in (\ref{schrodinger}),  from which we find that 
\beq\label{energy}
z = z(k) = -2t \cos (ka)\,.
\eeq 
Next,  we eliminate $\psi_0 = A = {t'\over t} e^{ika}\psi_1$ from the $n=0$ equation, and substitute it in the $n=1$ equation. In this way  we discover that 
\beq\label{1site}
-{t'^2\over t} e^{ika}\psi_1  + (H\vec\psi)_1 = z(k) \psi_1\,.
\eeq
The remaining equations $(H\vec\psi)_n = z(k) \psi_n\,\, (n\geq 2)$, as well as the equation (\ref{1site}) for $n=1$, can be neatly written as 
\beq\label{effective1}
\left(H - {t'^2\over t} e^{ika} P\right) \vec\psi  = z(k) \vec\psi\,,
\eeq
where $P=|1\rangle\langle 1|$ is the projector on site $1$, to which the lead is connected. 
Thus, we can ignore the lead from now on, and describe the open system itself by an {\em effective} hamiltonian 
\beq\label{effective}
H_{eff}  = H - \eta e^{ika} P\,,
\eeq
with 
\beq\label{eta}
\eta = {t'^2\over t}\,.
\eeq
$H_{eff}$ is a non-hermitian $N\times N$ matrix, which also depends on energy, through $k$, according to (\ref{energy}). This explicit $k$ dependence is of course, the price we had to pay in order to eliminate the lead from the description of  our system, and it should actually be expected of 
an effective description - the effective hamiltonian normally depends on the energy scale one studies.  Of course, when $t'=\eta=0$, the lead is disconnected, and $H_{eff} = H$ 
of the closed system. 

The picture we have in mind is that starting with the closed system ($\eta=0$), all $N$ energy eigenstates are real and sharply defined. Then we open the system adiabatically, 
i.e., increase $\eta$ slowly. As a result, each sharply defined energy eigenstate in the original system should broaden continuously into a resonance, with complex energy $z(k,\eta)$. Thus, we end up with $N$ complex resonance energies, the solutions of $H_{eff}(k) \vec\psi = -2t\cos(ka)\,\vec\psi$.  

Let us briefly elaborate on the domain in the complex-$k$ plane which corresponds to resonances. We shall follow the discussion in \cite{hatano, hf}, which give probabilistic interpretation to the modulus square of the time-dependent resonance eigenfunction $\Psi_n(t)$.  For this, we shall temporarily include the lead in our discussion. 
In a resonance (i.e., quasi-stationary) state the amplitude $\Psi_n(t)  = \psi_n e^{-i z(k) t} $ has to grow in magnitude into the lead, where the particle is likely to be after a long time. For the same reason, it also has to decay as function of time, at any finite fixed site.  Thus, in addition to (\ref{leftmoving}), we must also demand that both ${\rm Im}\,k$ and ${\rm Im}\, z = 2t\sin ({\rm Re}\,ka)\cdot\sinh ({\rm Im}\,ka)$ be {\em negative}. Thus, $\sin ({\rm Re}\,ka)\ >0$, which holds automatically
due to (\ref{leftmoving}). As was originally discussed in \cite{hatano} (and later extended in \cite{hf}), we see that as time goes by, we can maintain the numerical value of the spatial integral of  $|\Psi_n(t)|^2$ (the probability) ,  provided we allow the integration domain to expand at  a constant speed to the left, which is nothing but the ballistic velocity of the ejected particle. To summarize, resonances must all lie in the strip
\beq\label{resonance}
0\leq{\rm Re}\,ka\leq\pi\,,\quad {\rm Im}\,k <0\,
\eeq
in the fourth quadrant of the complex $k$ plane. Similarly, anti-resonances, which describe a situation in which the system absorbs particles from the lead\footnote{Here
the wave should propagate freely in the lead to the right, towards the system. Hence $-\pi\leq {\rm Re}\,ka \leq 0$. Furthermore,   ${\rm Im}\,k <0$, since at $t=0$ the it is overwhelmingly probable to find the particle in the lead, while at the same time we must also have   ${\rm Im}\, z >0$, since the probability to find the particle at $n=0$ must grow.}, must all lie in the strip $-\pi\leq{\rm Re}\,ka\leq 0\,,\quad {\rm Im}\,k <0\,$ in the third quadrant of the complex $k$ plane.

\subsection{The Secular Equation}
Resonances are the roots of the equation
\beq\label{secular}
\det\left(z  - H_{eff}\right)  = 0\,,
\eeq
with $k$ lying in the appropriate strip (\ref{resonance}). Practically, it is easier to compute the ratio of determinants
\beq\label{ratio}
F(k)  = {\det \left(z-H_{eff}\right)\over \det \left(z-H\right)}  =  {\det \left(z-H + \eta e^{ika} P\right)\over \det \left(z-H\right)}  = \det\left(1 + \eta e^{ika} GP\right)\,,
\eeq
where 
\beq\label{green}
G = {1\over z-H}
\eeq
is the resolvent of $H$.  In (\ref{secular})-(\ref{green}) we must of course set $z$ according to (\ref{energy}). 

Note that  $(GP)_{nm} = G_{n1}\delta_{m1}$. Hence, $1 + \eta e^{ika} GP$ is a lower diagonal matrix, and computation of  the last determinant in 
(\ref{ratio}) is immediate. We find simply that 
\beq\label{Fk}
F(k) = 1 + \eta e^{ika} G_{11}\left(z(k)\right)\,.
\eeq
Thus, in order to solve for the resonance spectrum of our model, all we require is the $G_{11}$ element of the Green's function of the original closed system. 
The latter is the Green's function of a hermitian hamiltonian,  and therefore well-studied. Note that we have not specified the specific nature of the closed 
system corresponding to $H$. Our discussion is completely generic!

\subsection{The DOR}
For a given realization of $H$, $F(k)$ is a holomorphic function of $k$, and has zeros at the eigenvalues of $H_{eff}$ and poles (on the real axis) at the eigenvalues of $H$. 
Let $k_\alpha^0$ and $k_\beta^p$ be, respectively, the zeros and (purely real) poles of $F(k)$. Thus,  
\beq\label{logder}
{F'(k)\over F(k)}  = \sum_\alpha{1\over k - k_\alpha^0}  - \sum_\beta{1\over k - k_\beta^p} \,.
\eeq
From the identity 
\beq\label{Gauss}
{\partial\over\partial k^*} {1\over k-q}  = \pi\delta^{(2)}(k-q)\,,
\eeq
which is nothing but Gauss' Law in 2d electrostatics (for a unit point charge located at position $k=q$), we thus find 
\beq\label{preDOR}
{1\over \pi}{\partial\over\partial k^*}{F'(k)\over F(k)} =  \sum_\alpha \delta^{(2)}( k - k_\alpha^0)  - \sum_\beta\delta^{(2)}(k - k_\beta^p) \,.
\eeq
Averaging this equation with its complex-conjugate, we finally obtain that 
\beq\label{DOR}
\rho(k,k^*) = {1\over 2\pi}{\partial^2\over\partial k\partial k^*}\log\Big|F(k)\Big|^2 =  \sum_\alpha \delta^{(2)}( k - k_\alpha^0)  - \sum_\beta\delta^{(2)}(k - k_\beta^p) \,.
\eeq
Since the poles live entirely on the real axis, going off it and into the fourth quadrant in the complex $k$-plane, we obtain our desired DOR.

Continuing the analogy with 2d electrostatics \cite{electrostatics}, observe that (\ref{DOR}) is nothing but the Poisson equation, relating the charge distribution on the LHS, to the electrostatic potential
\beq\label{electrostaticpot}
W(k,k^*) = -\log\Big|F(k)\Big|^2 = -\log{\Big|\det(z-H_{eff})\Big|^2\over\Big|\det(z-H)\Big|^2}
\eeq
on the RHS. Moreover, note that the real quantity $\Big|\det(z-H_{eff})\Big|^2$ in (\ref{electrostaticpot}) is proportional to the determinant of the $2N\times 2N$ {\em hermitian}
operator
\beq\label{hermitized}
{\cal H} = \left(\begin{array}{cc} 0 & z-H_{eff}\\{}&{}\\
z^*-H_{eff}^\dagger & 0\end{array}\right)\,.
\eeq
In fact, given a non-hermitian operator, such as $H_{eff}$,  whose spectrum we wish to study, the method of hermitization \cite{FZ} instructs us to construct its hermitized form
(\ref{hermitized}), and study its spectrum, which of course lies entirely on the real axis. Thus, for example, the Green's function $1/(\zeta -{\cal H})$ is analytic in the complex $\zeta$ plane, save for poles (or a cut, upon averaging) along the real axis, where the spectrum is located. Thus, one may bring the power of analytic function theory to bear in analyzing the spectrum, which cannot be done for the non-hermitian $H_{eff}$.  

\subsection{The Averaged DOR}
In our closed disordered system
\beq\label{G11}
G_{11}(z) = X(z) + iY(z)
\eeq
is a complex valued random variable, with probability distribution
\beq\label{probG}
{\cal P}(X,Y;z) =  \langle \delta\left(X-X(z)\right)\delta\left(Y-Y(z)\right)\rangle\,,
\eeq
which we assume to be known. Thus, from (\ref{DOR}) and (\ref{probG}) we immediately obtain an integral representation for the averaged DOR as 
\beq\label{aveDOR}
\rho_{av}(k,k^*) = {1\over 2\pi}{\partial^2\over\partial k\partial k^*}\,\int\,dX dY {\cal P}(X,Y;z(k))\log\Big|1 + \eta e^{ika} (X + iY)\Big|^2\,.
\eeq

\section{The One-Dimensional Disordered Chain and the Continuum Limit of its $H_{eff}$}
We shall now depart from the general discussion and take $H$ to be the tight-binding hamiltonian of a disordered chain with $N$ sites, i.e., the one-dimensional Anderson model. We take the nearest-neighbor hopping amplitudes to be $t$, as in the lead. The site energies $\epsilon_n\,\, (n = 1,2,\ldots,N)$ are i.i.d. random variables taken from some 
probability distribution $q(\epsilon)$. Thus, the corresponding hermitian matrix $H$  in (\ref{schrodinger}) (and in Fig.\ref{lead}), in the previous section, is given by 
\beq\label{HAnderson}
H_{mn} = -t(\delta_{m,n+1} + \delta_{m+1,n}) + \epsilon_n\delta_{mn}\,,\quad 1\leq m,n \leq N\,.
\eeq
The resulting Schr\"odinger equation is therefore 
\beq\label{chain}
-t(\psi_{n+1} + \psi_{n-1}) + \epsilon_n\psi_n = z\psi_n\,,
\eeq
with Dirichlet boundary conditions 
\beq\label{dirichlet} 
\psi_0=\psi_{N+1} = 0\,,
\eeq
corresponding to a {\em closed chain}. As can be seen from (\ref{effective}), the effective 
Schr\"odinger equation for the {\em open system} (with the lead eliminated, of course) is obtained from (\ref{chain}) (or (\ref{HAnderson})) simply by replacing $\epsilon_n$ by 
\beq\label{tilde-epsilon}
\tilde\epsilon_n   = \epsilon_n - \eta e^{ika}\delta_{n1}\,.
\eeq
Statistics of resonances in this model was studied in detail in \cite{ks1, ks2, terraneo}.

The effective Schr\"odinger equation  $(H_{eff}-z)\vec\psi=0$ can be formally obtained by applying the variational principle 
\beq\label{variational}
{\delta S\over\delta \vec\psi^\dagger} = (H_{eff}-z)\vec\psi = 0
\eeq
to the complex action 
\beq\label{action}
S = \sum_{n=1}^N a\left[(\epsilon_n-\zeta )|\psi_n|^2 - D{\delta_{n1}\over a}\left(\psi_1^*{\psi_2-\psi_1\over a} + {\left({t'\over t}\right)^2 e^{ika} -1\over a}|\psi_1|^2\right)\right]
-D\sum_{n=2}^N a\psi_n^*{\delta^2\psi_n\over a^2}\,,
\eeq
where
\beq\label{def1}
D = ta^2\quad {\rm and}\quad \zeta = z+ 2t = 4t\sin^2\left({ka\over 2}\right)
\eeq
are, respectively, the diffusion constant (the lattice version of ${\hbar^2\over 2m}$) and the shifted (renormalized) energy, 
and 
\beq\label{difference}
\delta^2\psi_n = \psi_{n+1} - 2\psi_n + \psi_{n-1}
\eeq
is the symmetric second difference (the lattice discretized version of the self-adjoint laplacian).  
 
Strictly speaking, we should really apply the variational principle to the {\em real} action corresponding to the hermitized form (\ref{hermitized}).  However, in order to keep the discussion as brief as possible, and since all we want to obtain in this section is the continuum limit of $H_{eff}$, and not to pursue the averaged DOR in detail, we shall contend ourselves with the complex action $S$. 

\subsection{The Continuum Limit}
The continuum limit is obtained by sending $a\rightarrow 0$ and $t\rightarrow \infty$ simultaneously, while keeping $D = ta^2 =  {\hbar^2\over 2m} $ and $k$ finite. 
Furthermore, $t'\rightarrow\infty$ as well, such that the ratio 
\beq\label{lambdaratio}
\left({t'\over t}\right)^2 = e^{\lambda a}
\eeq
with $\lambda$ finite. 
In this limit we also obtain the familiar relation $\zeta   =  Dk^2$. 
In the limit, the lattice amplitudes tend to the continuous wave function,  $\psi_n = \psi(na)\rightarrow\psi(x)$ and the site energies tend to the potential $\epsilon_n\rightarrow V(x)$ . Obviously, ${\delta^2\psi_n\over a^2}\rightarrow\partial_x^2\psi(x)$ 
and ${\psi_2-\psi_1\over a} \rightarrow \psi'(a+)$. Finally, of course, ${\delta_{n1}\over a}\rightarrow \delta(x-a)$ and $\sum_n a\rightarrow\int\,dx$, and (\ref{dirichlet}) tend to the continuum Dirichlet boundary conditions
\beq\label{cont-dirichlet}
\psi(0) = \psi(L) = 0
\eeq
with $L=Na$ (where of course $N\rightarrow\infty$). Plugging all these limiting quantities in (\ref{action}), we obtain the continuum limit of $S$ as
\beqra\label{cont-action}
S &=& \int\limits_{0+}^L \,dx\,\left[\left(V(x) - \zeta\right)|\psi(x)|^2 -D\psi^*(x)\partial_x^2\psi(x) - D\delta(x-a)\left(\psi^*(x)\partial_x\psi(x) + (\lambda + ik)|\psi(x)|^2\right)   \right] \nonumber\\{}\nonumber\\
&=&
 \int\limits_{0+}^L \,dx\,\psi^*(x)\left(H_{eff}^{cont} - \zeta\right)\psi(x)\,,
 \eeqra
with the boundary conditions (\ref{cont-dirichlet}) understood. We can immediately read off the continuum effective effective hamiltonian from (\ref{cont-action}), namely, 
\beq\label{continuum-effective}
H_{eff}^{cont}  = Dp^2 + V(x) -D\delta(x-a)\left(ip + \lambda + ik\right) 
\eeq
with $p= -i\partial_x$ as usual. 

Note that we have explicitly left the infinitesimal lattice spacing $a$ in (\ref{cont-action}) and (\ref{continuum-effective}) as a mnemonic. In fact,  $\delta(x-a)$ in these expressions really stands for a thin boundary layer around the left end of the chain, with a very large coefficient, which penalizes for having $\partial_x\psi(x) + (\lambda + ik)\psi(x) \neq 0 $ in the immediate vicinity of $x=0+$. The continuum Schr\"odinger equation derived by applying the variational principle to (\ref{cont-action}) generates in this way the continuum {\em resonance boundary condition} 
\beq\label{resonance-bc} 
\partial_x\psi(0+) + (\lambda + ik)\psi(0) = 0\,,
\eeq
which depends on energy, through $k$. Since in the lead, $x<0$ we have the outgoing wave $\psi(x) = \psi(0) e^{-ikx}$,  the derivative $\partial_x\psi(x)$ jumps:
$\psi'(0+) - \psi(0-)  = -\lambda\psi(0)$. This jump is the result of the singular contact potential term $-\lambda D\delta(x)$ in (\ref{continuum-effective}).   
Note from (\ref{continuum-effective}) that $\lambda\rightarrow -\infty$ penalizes for having $\psi(0)\neq 0$. Thus, this limit corresponds to Dirichlet boundary conditions, namely, $t'=0$ and a closed chain, as can be seen also from (\ref{lambdaratio}).

More precisely, for $a$ very small but finite, one integrates the Schr\"odinger equation $-D\partial_x^2\psi(x)  + V(x)\psi(x) = \zeta\psi(x)$, subjected to $\psi(L) = 0$, from the right end of the system all the way to $x=a$, where the large coefficient of the boundary layer interaction takes over, and fixes $\psi'(a-) = -(\lambda + ik)\psi(a). $  The wave function has then to relax to zero at $x=0$ across the thin boundary layer, with tremendous slope. This segment of the wave function is an artifact, which we cut and throw, and replace by the resonance boundary condition (\ref{resonance-bc}). 

\section{Concluding Remarks Concerning Higher Dimensional Systems}
The structure revealed by analyzing the continuum limit of the one dimensional case is quite generic. The effective non-hermitian hamiltonian 
is generically given by the original differential expression (in the coordinate representation), supplemented by an appropriate energy dependent complex boundary condition. 
In conclusion, let us mention briefly two simple 3d examples in the continuum, which correspond to having {\em infinitely} many weak channels connecting the system to the environment. These results are straightforward: 
\begin{itemize}
\item{Disordered half-space coupled uniformly to the environment  through a contact plane.}\\ 
Let's take the disordered system to live in the $z>0$ half-space, and let it communicate with its environment through the $xy$ plane. 
For a given complex energy $z = {\hbar^2Q^2\over 2m}$ there is a continuum of resonances indexed by the components of the wave-vector ${\bf k}_\perp$, perpendicular to the $z$ axis, which are real. They correspond to the direction in which the particle is ejected from the system. Let $q=k_3$ be the on-shell complex component of momentum in the $z$-direction, such that $q^2 = Q^2 - k_\perp^2$. Then, the resonance amplitude
{\em immediately outside} the system, must satisfy the outgoing boundary condition
\beq\label{plate}
(\partial_z + iq\,\sign \,{\rm Im}\,q)\psi_{\bf k_\perp}(z=0^-)  = 0\,. 
\eeq
We can then obtain $\partial_z\psi_{\bf k_\perp}(z=0^+)$ right inside the system by considering the contact potential $-D\lambda\delta(z)$, in complete analogy with (\ref{resonance-bc}).  Since the boundary condition (\ref{plate}) is rotationally symmetric with respect to the $z$-axis, we expect the {\em averaged} DOR to inherit this symmetry as well.   
\item{A disordered ball of radius $a$ coupled uniformly to the environment through its surface.}\\
In this case, we should consider resonances with definite angular momentum quantum numbers $l,m$ in the outside world. 
For a given complex energy $z = {\hbar^2Q^2\over 2m}$, the corresponding outgoing wave amplitude $\psi_{lm}(r) = A_{lm} h_l(Qr)$ must be proportional to a Hankel function. Thus, it must trivially satisfy  
\beq\label{sphere}
\psi_{lm}'(a+)  = Q{h_l'(u)\over h_l(u)} \psi_{lm}(a)\,,
\eeq
where $u=Qa$. Again, the radial derivative immediately inside the ball may be obtained by taking into account the jump in the radial derivative due to a uniform radial-shell contact potential. Due to spherical symmetry, the boundary condition (\ref{sphere}) is independent of $m$. Consequently, the averaged DOR should inherit this property as well. 
\end{itemize}

\vspace{1cm}
{\bf acknowledgements}~ I wish to thank Boris Shapiro for many valuable discussion on resonances in disordered systems. 
This work was supported in part by the Israel Science Foundation (ISF). 

\end{document}